\documentclass[10pt,twocolumn,aps,english,pra,superscriptaddress,longbibliography]{revtex4-2} 

\usepackage[utf8]{inputenc}

\usepackage{tikz} % Package for drawing
\usepackage{amsmath,amsfonts,amsthm, mathtools}
\usepackage{placeins}
\usepackage{float}
\usepackage{dsfont}
\usepackage{csquotes}
\usepackage{amssymb}
\usepackage{verbatim}
\usepackage{bm}
\usepackage{amsmath}
\usepackage{amssymb}
\usepackage{stmaryrd}
\usepackage{amsthm}
\usepackage[caption=false]{subfig}
\usepackage{physics}
\usepackage{bm}  

\usepackage{hyperref}
\definecolor{darkred}  {rgb}{0.5,0,0}
\definecolor{darkblue} {rgb}{0,0,0.5}
\definecolor{darkgreen}{rgb}{0,0.5,0}
\hypersetup{
  colorlinks = true,
  urlcolor  = blue,         % color of external links
  linkcolor = darkblue,     % color of internal links
  citecolor = darkgreen,    % color of links to bibliography
  filecolor = darkred       % color of file links
}

\theoremstyle{definition}

\newtheorem{thm}{Theorem}

\newcommand{\mbf}{\mathbf}
\newcommand{\mbb}{\mathbb}

\usepackage{multirow}

\definecolor{cool_green}{rgb}{0.0, 0.5, 0.0}

\begin{document}

\title{Random Distillation Protocols in Long Baseline Telescopy }
\author{Yunkai Wang}
\email{yunkaiw2@illinois.edu}
\affiliation{IQUIST, University of Illinois at Urbana-Champaign, Urbana, IL 61801 USA}
\affiliation{Department of Physics, University of Illinois at Urbana-Champaign, Urbana, IL 61801 USA}

\author{Eric Chitambar }
\email{echitamb@illinois.edu}
\affiliation{IQUIST, University of Illinois at Urbana-Champaign, Urbana, IL 61801 USA}
\affiliation{Department of Electrical and Computer Engineering, University of Illinois at Urbana-Champaign, Urbana, IL 61801 USA}

\begin{abstract}

In quantum-enhanced astronomical imaging, multiple distant apertures work together by utilizing quantum resources distributed from a central server. Our findings suggest that pre-processing the stellar light received by all telescopes can improve imaging performance without increasing resource consumption.
The pre-processing leverages weak quantum measurements and modifies random-party entanglement distillation protocols from quantum information science. Intuitively, this approach allows us to collapse the stellar light that is originally coherent between all telescopes to one pair of telescopes with probability arbitrarily close to one. 
The central server can then distribute entanglement solely to the pair of telescopes receiving a photon, thereby enhancing the efficiency of resource utilization. We discuss two types of resources that benefit from this pre-processing: shared entanglement and a shared reference frame.
\end{abstract}

\maketitle

\textit{Introduction} - 
Interferometric imaging is a powerful technique in astronomy that combines multiple telescopes to simulate a larger one \cite{monnier2003optical,rohlfs2013tools}. Based on the van Cittert-Zernike theorem \cite{zernike1938concept}, it relates the mutual coherence function of light between distant telescopes to a Fourier component of the source's intensity distribution, determined by the telescope baseline. By measuring different baselines, the intensity distribution can be reconstructed, enabling state-of-the-art resolution in astronomy. This method, for instance, was used to capture the first image of a supermassive black hole in Messier 87 \cite{event2019first}.

%Interferometric imaging is a well-established and highly effective technique in astronomy that leverages multiple telescopes to effectively act as one larger telescope  \cite{monnier2003optical,rohlfs2013tools}. Interferometric imaging is based on the van Cittert-Zernike theorem \cite{zernike1938concept}, which states that the mutual coherence function between light received by distant telescopes corresponds to a Fourier component of the source's intensity distribution. The associated spatial frequency is determined by the baseline between the telescopes. By measuring multiple Fourier components using different baselines, one can reconstruct the intensity distribution of the source. Interferometric imaging provides the state-of-the-art resolution in astronomical observations. For example, this advanced technique enabled the first image of a supermassive black hole at the center of the Messier 87 Galaxy \cite{event2019first}.

Measuring the coherence function typically requires some resources distributed by a center server, and the large distances between telescopes can pose challenges for resource distribution.  We will consider two significant types of resources in this context. The first type of resource is entanglement.
Traditionally, astronomical interferometers bring light from distant telescopes to a central point for interference, which leads to transmission loss. It has been proposed that distributing entanglement between telescopes allows for nonlocal measurements via teleportation, avoiding transmission loss \cite{gottesman2012longer}. Follow-up studies have investigated ways to reduce entanglement consumption using quantum memories \cite{khabiboulline2019optical, khabiboulline2019quantum} and enhance the protocol with controlled quantum gates \cite{czupryniak2023optimal}. Additional studies also examined entanglement costs \cite{czupryniak2022quantum}, the benefit of using multiple entangled states without quantum repeaters \cite{marchese2023large}, and a continuous-variable version of entanglement-based interferometry \cite{huang2024limited, wang2023astronomical}.

%Traditionally, astronomical interferometers bring light from distant telescopes to a single location to interfere and measure the coherence function, a process that suffers from transmission loss. It has been proposed that distributing entanglement to distant telescopes can avoid transmission loss \cite{gottesman2012longer} since the shared entanglement essentially allows the distant telescopes to perform a nonlocal measurement through teleportation.   This idea has been further explored in several subsequent studies. Some works suggested reducing the amount of consumed entanglement with the help of quantum memories at telescopes \cite{khabiboulline2019optical, khabiboulline2019quantum}, while others proposed using controlled quantum gates to improve the original proposals \cite{czupryniak2023optimal}. The entanglement cost in quantum network-assisted telescopes has been analyzed \cite{czupryniak2022quantum}.  Additionally, possible quantum enhancements gained by distributing multiple copies of entangled states without quantum repeaters \cite{marchese2023large}, and a continuous-variable version of the astronomical interferometer assisted by entanglement have been proposed \cite{huang2024limited, wang2023astronomical}. 

The second type of resource is a shared reference frame.  In entanglement-based quantum telescopy just described, distributed entanglement  also serves as a shared frame, enabling coherence function measurements that respect the photon-number superselection rule \cite{bartlett2007reference}. Without entanglement, only local measurements are possible, and measuring the coherence function implicitly requires correlating local reference frames at each telescope. Synchronizing atomic clocks at each telescope, the standard approach in radio interferometry \cite{wilson2009tools}, can establish a shared reference, though these clocks lack the precision for optical frequencies. Alternative methods for sharing a reference frame in the optical domain require distributing states of physical systems that get consumed in the implementation of local measurements.  For example, performing heterodyne detection between two distant telescopes requires phase-locked lasers at each telescope that interfere with stellar light \cite{hale2000berkeley}.

In this work, we consider a network of $M$ telescopes with varying distances that provide different baselines for stellar interferometry. We focus on the simplest case where the central server distributes only bipartite resources—either entanglement or a shared reference frame—to one  pair of telescopes per time step. While distributing multipartite resources, like tripartite entanglement or multi-party phase locking, could be explored, it is experimentally more challenging. So, here we just focus on bipartite resources; a consideration of multipartite resources is discussed in the conclusion of this work. For bipartite protocols, the question then becomes to which pair of telescopes should the central server distribute resources at each time step.  

%In this work, we consider a network of $M$ telescopes, and the differing spatial distance between pairs of telescopes provides differing baselines for stellar interferometry.  We restrict to the simplest scenario in which the central server is able to distribute just bipartite resources between the telescopes.  That is, in each time step of the protocol the server distributes either bipartite entanglement or a shared reference frame to some pair of telescopes.  Of course, one could consider the distribution of, say, tripartite entanglement or multi-party phase locking.  However, experimentally this can be much more demanding, and so here we just focus on bipartite resources; a consideration of multipartite resources is discussed in the conclusion of this work.  For bipartite protocols, the question then becomes to which pair of telescopes should the central server distribute resources at each time step.  

Assuming the amplitude of a single photon in the stellar light is uniform at each telescope, then the probability of detecting a stellar photon at any given telescope is $\frac{1}{M}$, provided a single photon is detected at one of them.  Hence, without any further information, the central server's best strategy is to just flip a random coin and distribute the bipartite resources to one of the $\binom{M}{2}$ different pairs of telescopes.  If the stellar photon is not detected at one of the $(M-2)$ unselected telescopes, then the interfometric measurement can be performed by the pair of selected telescopes.  We will refer to this random coin selection process as quantum telescopy with \textit{classical randomness}.  Clearly, this classical randomness strategy seems wasteful since it requires the server to coincidentally select the ``right'' pair of telescopes; i.e. the pair that detects the stellar photon. 

\begin{figure}[t]
\centering
\includegraphics[width=0.8\columnwidth]{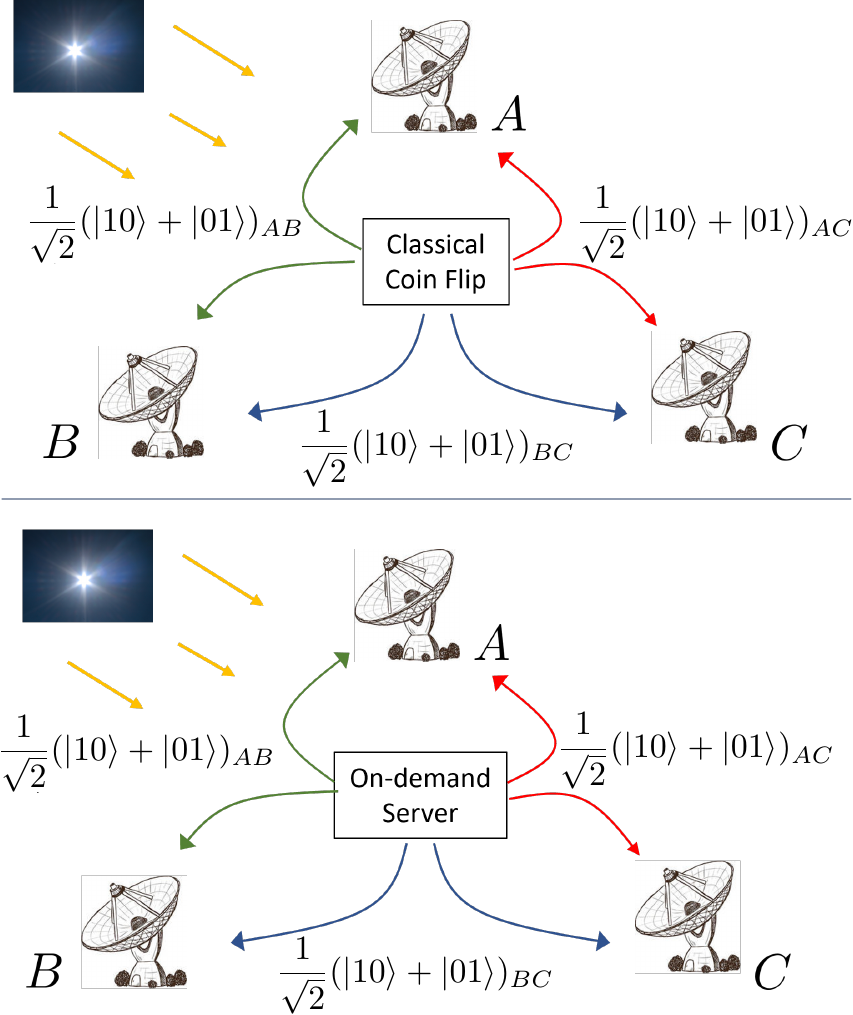}
\caption{We compare two methods for quantum long baseline telescopy. In the top method, a source distributes entanglement to a random pair of telescopes so that it can interfere with light from a distant star.  In the bottom method, the stellar photon undergoes pre-processing to localize the photon to a pair of telescopes and then the source sends entanglement to that pair. } 
\label{stellar_randomness}
\end{figure}

The central goal of this letter is to investigate an improved method of resource distribution that involves a pre-processing of the stellar photon.  Instead of randomness from a classical coin, our proposed protocol uses the intrinsic randomness of quantum measurements in a pre-processing phase.  The idea is an adaptation of the random-party entanglement distillation protocol studied in Refs. \cite{fortescue2007random,fortescue2008random,cui2011randomly} that converts a W-class state into an Einstein-Podolsky-Rosen (EPR) pair shared between a random pair of parties.   Applying random-party distillation on a stellar photon before subjecting it to an interference measurement collapses its state, which is originally coherent across all telescopes, to a single pair of telescopes. Once the photon is localized to a known pair of telescopes, the central server can then distribute the resources to this specific pair (see Fig. \ref{stellar_randomness}), and overall we achieve enhanced performance using the same amount of resources.  The rest of this letter is devoted to quantifying how much of an improvement this pre-processing provides in terms of the Fisher information per photon distributed by the central server.  For entanglement-based interferometric protocols we find that the relative advantage scales linearly in $M$ while for local measurement schemes it scales quadratically.

%We explore the following question: if we pre-process the stellar light, can we enhance the performance of measurements while consuming the same amount of resources distributed to distant telescopes? We find that the  protocol proposed to distill the W-state into Einstein-Podolsky-Rosen (EPR) pairs in Ref. \cite{fortescue2007random} can be adapted to pre-process the stellar light. This pre-processing procedure collapses the stellar states, which are originally coherent across all telescopes, to a single pair of telescopes. By distributing resources to this pair of telescopes, we achieve enhanced performance using the same amount of resources.

%Intuitively, since we reduce the resource consumption of $M$ telescopes to the consumption of $2$ telescopes, we save the distributed resources by a factor of $M/2$. 

\textit{Entanglement and random-party distillation} - We begin with entanglement-based astronomical imaging and describe the idea of random-party pre-processing of the stellar photon.  We first derive the results for three telescopes and then generalize to $M$ telescopes.  As shown in Fig.~\ref{stellar_randomness}, three telescopes $A,B,C$  receive light from a distant stellar source described by the density matrix \cite{mandel1995optical}
\begin{equation}
    \rho_{ABC} = (1 - \epsilon) |000 \rangle \langle 000| +\epsilon \rho^{(1)}_{ABC}+O(\epsilon^2),
\end{equation}
\begin{equation}
  \hspace{-2cm}\text{where}\qquad  \rho_{ABC}^{(1)} = \frac{1}{3} \begin{pmatrix}
    1 & g_{AB} & g_{AC} \\
    g_{AB}^* & 1 & g_{BC} \\
    g_{AC}^* & g_{BC}^* & 1\\
    \end{pmatrix},
\end{equation}
and $\epsilon\ll1$ is the mean photon number per temporal mode, which is usually much less than one in the optical wavelength.  The coherence term is given by $g_{XY}=a_{XY}+ib_{XY}$, where  $XY\in\{AB,AC,BC\}$.   The density matrix $\rho^{(1)}_{ABC}$ is written in the single photon basis $\ket{100},\ket{010},\ket{001}$ at each telescopes; i.e. $\ket{1}_X$ is a single photon at telescope $X$ and $\ket{0}_X$ is the vacuum state. %Our goal is to estimate each of the coherence function $g_{XY}$ by measuring $\rho_{ABC}$. 

The seminal proposal by Gottesman \textit{et al.} \cite{gottesman2012longer} involves a central server distributing a dual-rail photon $\frac{1}{\sqrt{2}}(\ket{01}+\ket{10})_{XY}$ to one pair of telescopes $X,Y$. This terrestrial photon then interferes with the stellar light at telescopes $X,Y$ to extract information about $g_{XY}$. This scheme will be referred as the Gottesman-Jennewein-Croke (GJC) protocol in our work.  Note that dual-rail single-photon states typically require additional reference-frame states to function as one entangled bit (ebit) in most quantum information tasks \cite{tan1991nonlocality,bartlett2007reference}. However, in the GJC protocol, the stellar photon can naturally serve as the reference frame for terrestrial photons, making it appropriate to treat dual-rail single-photon states as entanglement resources.  We consider the Fisher information $F$ in estimating each of the coherence functions $\{g_{AB}, g_{AC}, g_{BC}\}$ \textit{per dual-rail terrestrial photon generated}, whose inverse lower bounds the variance $V$ of estimating these unknown classical parameters, i.e. $V\geq F^{-1}$.

For the GJC protocol with \textit{classical randomness}, the server determines which pair of parties should receive the photon by flipping a uniform classical coin.  If parties $XY$ receive the terrestrial photon, then the dual-rail structure allows $XY$ to realize a nonlocal measurement on $\rho_{ABC}$ described by a positive operator-valued measure (POVM) whose elements are computed in Ref.~\cite{tsang2011quantum}.  If we incorporate the classical randomness as part of the measurement, then the  entire process can be described by a single positive operator-valued measure (POVM).  The elements of this POVM that contribute to the Fisher information have the form:
\begin{align}
   & E^{\pm}_{XY}= \frac{1}{2\binom{M}{2}}\op{\delta_\pm}{\delta_\pm}_{XY}\otimes \mbb{I}_{\overline{XY}},  
\end{align}
where $M=3$ for the three telescope case, $XY\in\{AB,AC,BC\}$,  $\overline{XY}$ is the complement of set $XY$, and $\ket{\delta_{\pm}}_{XY}=\frac{1}{\sqrt{2}}(\ket{01}\pm e^{i\delta}\ket{10})$.  The factor of two in the denominator arises because the GJC POVM succeeds only with probability one-half, while the factor of $\binom{M}{2}$ accounts for the fact that the server distributes resources to $\binom{M}{2}$ possible pairs of telescopes.  %In this work, for convenience, we will write the system labels $XY$ as subscripts for states and as superscripts for POVMs. 
As detailed in Sec.~\ref{SI:POVM_gottesman_classical} of the Supplemental Material, the Fisher information $F$ for estimating the $\{g_{AB},g_{AC},g_{BC}\}$ using this POVM decomposes into three $2\times 2$ blocks of the form 
\begin{align}\label{classical_three}
    F_{XY}^{cr}=\frac{\epsilon}{9\left[1-Re(g_{XY}e^{-i\delta})^2\right]}\begin{pmatrix}\cos^2\delta&\sin\delta\cos\delta\\\sin\delta\cos\delta&\sin^2\delta\end{pmatrix};
\end{align}
i.e. $F=F_{AB}\oplus F_{AC}\oplus F_{BC}$. Note that Eq.~\ref{classical_three} is the Fisher information matrix \textit{per terrestrial photon}.

%\subsubsection{Gottesman Protocol with Quantum Randomness}\label{three telescopes quantum}
In the GJC protocol with \textit{quantum randomness}, the stellar light is pre-processed before distributing the terrestrial photon. This approach allows us to leverage the intrinsic randomness of quantum measurement to decide which pair of telescopes receives the terrestrial photon, instead of using a classical coin. The protocol is based on the random-party entanglement distillation protocol proposed by Fortescue and Lo \cite{fortescue2007random}. On the stellar state $\rho_{ABC}$, each party performs a local weak quantum measurement described by Kraus operators expressed in the photon number basis as 
\begin{align}\label{eq:weak_measurement}
    M_0&=\begin{pmatrix}\sqrt{1-\tau}&0\\0&1\end{pmatrix}&M_1&=\begin{pmatrix}\sqrt{\tau}&0\\0&0\end{pmatrix}.
\end{align}
They all classically announce their measurement outcomes.  If exactly two of the parties obtain outcome $M_0$ and the other party obtains outcome $M_1$, then the two parties getting $M_0$ will end up remaining coherently correlated while the third party is uncorrelated. The post-measurement state is given by $\rho_{ABC}=\rho_{AB}\otimes\ket{0}\bra{0}_C$. When this occurs, the central server distributes the dual-rail qubit to the entangled parties and they perform the two-party GJC measurement.  On the other hand, if all three parties obtain outcome $M_0$,  the post-measurement state is given by $\rho_{ABC} =  \frac{1}{P}\left((1-\tau)^3(1 - \epsilon) |000 \rangle \langle 000| + (1-\tau)^2\epsilon\rho^{(1)}_{ABC} \right)$,
where $P$ normalizes the state. The additional prefactor of the vacuum term is smaller than that of the single-photon term, resulting in a reduced fraction of vacuum in the state $\rho_{ABC}$, while the state $\rho^{(1)}_{ABC}$ remains unchanged.  In this case, all the parties perform the $\{M_0,M_1\}$ measurement again, and this process continues.  A failure occurs if at anytime during the protocol at least two parties obtain an outcome $M_1$. See Sec.~\ref{implementation_POVM} of the Supplemental Material for details on the post-measurement states and the implementation of this weak measurement. 

Suppose the parties agree in advance to perform $D$ total rounds of measurement. If only party $\overline{XY}$ obtains outcome $M_1$ in some round $r\in\{1,\cdots,D\}$ and the GJC protocol is then implemented on telescopes $XY$, the relevant POVM effect over all $D$ rounds is
\begin{align}
E_{XY}^\pm&= \frac{\gamma_D}{2}\op{\delta_\pm}{\delta_\pm}_{XY}\otimes\op{0}{0}_{\overline{XY}},
\end{align}
where $\gamma_D=\sum_{r=1}^D(1-\tau_r)\tau_r\left(\prod_{k=1}^{r-1}(1-\tau_{k})^2\right)$ for the three telescopes case, $\tau_r$ denote the strength of local measurement in the $r^{th}$ round of measurement.  We obtain the Fisher Information matrix \textit{per terrestrial photon}, which has the same $2\times 2$ blocks except with an overall multiplicative factor:
%\begin{align}\label{quantum_three}
    %F_{XY}^{qr}=\frac{\gamma_D\epsilon}{3\left[1-Re(g_{XY}e^{-i\delta})^2\right]}\begin{pmatrix}\cos^2\delta&\sin\delta\cos\delta\\\sin\delta\cos\delta&\sin^\delta\end{pmatrix}.
%\end{align}
\begin{align}\label{quantum_three}
F_{XY}^{qr}=3\gamma_D F_{XY}^{cr}.
\end{align}
If we choose $\tau_r=\frac{1}{2+D-r}$, then we find by induction that
$\gamma_D=\frac{D}{2(1+D)}$.
We can also use the optimized protocol \cite{Li-2020a}, which gives $\gamma_D=\frac{D+1}{2(D+2)}$, and modifies the Fortescue-Lo protocol by performing a hard measurement in the final round $D$.  More details about the derivation of $\gamma_D$ is given in Sec.~\ref{SI:POVM_gottesman_quantum} of the Supplemental Material. %If we let $F^{cr}$ and $ F^{qr}$ denote the Fisher Information matrix using classical randomness and quantum randomness derived in Eq.~\ref{classical_three} and Eq.~\ref{quantum_three} respectively, then
From Eq.~\eqref{quantum_three} we obtain
\begin{equation}
\frac{\Vert F^{qr}\Vert}{\Vert F^{cr}\Vert}=3\gamma_D \to \frac{3}{2}\quad\text{as $D\to\infty$}
\end{equation}
Even for $D=3$ (i.e. for three rounds of pre-processing) we obtain $\gamma_D>1/3$ and hence achieve a quantum enhancement for the Fisher information \textit{per terrestrial photon}.  With more telescopes, the effect becomes even more pronounced, as stated in the following theorem.
\begin{thm}
In the case of $M$ telescopes, the ratio of the Fisher information \textit{per terrestrial photon} for quantum randomness compared to classical randomness is given by
\begin{align}\label{ratio_entangle}
    \frac{\Vert F^{qr}\Vert}{\Vert F^{cr}\Vert}=\binom{M}{2}\gamma_D \to \frac{1}{M-1}\binom{M}{2}=\frac{M}{2}.
\end{align}
\end{thm}
%The theorem is proven in Sec.~\ref{SI:POVM_gottesman_quantum} of Supplemental. 

The numerical calculation shows that the optimal value for $\gamma_D$ is $1/(M-1)$ as shown in the Sec.~\ref{SI:POVM_gottesman_quantum} of Supplemental Material. Additionally, the Supplemental Material explores the optimal choice of $\tau_r$  in each round $r$.  However,  we find that  the ansatz of $\tau_r=\frac{1}{2+D-r}$ yields a ratio ${\Vert F^{qr}\Vert}/{\Vert F^{cr}\Vert}$ that still asymptotically approaches the same value $\binom{M}{2}\gamma_D$ as the numerically optimized choice of $\tau_r$.   Furthermore, it is evident that when $M=8$, a single round of measurement with an optimized choice of $\tau_r$ enables a ratio ${\Vert F^{qr}\Vert}/{\Vert F^{cr}\Vert}$ strictly larger than one. %Additionally, for $D \geq 3$, even with a suboptimal choice of $\tau_r=1/(2+D-r)$, the quantum randomness scheme still surpasses its classical counterpart. 
In summary, we have found that for any number of telescopes, at most a few rounds of pre-processing are needed to demonstrate an advantage over classical randomness, highlighting its potential for real-world applications.

\begin{figure}[!tbh]
\begin{center}
\includegraphics[width=0.8\columnwidth]{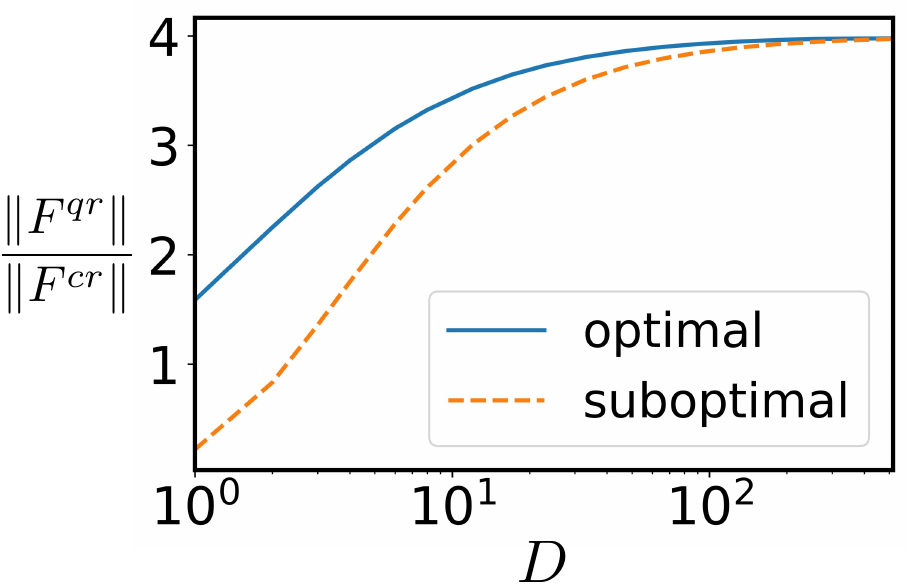}
\caption{Ratio between Fisher information  with quantum and classical randomness $\frac{\Vert F^{qr}\Vert}{\Vert F^{cr}\Vert}$ as a fuction of the number of rounds $D$. The solid line uses numerically optimized $\tau_j$, the dashed line uses suboptimal choice of $\tau_j = \frac{1}{2 + D - j}$ for $j$th round. The number of telescopes  $M = 8$. } 
\label{suboptimal_FI}
\end{center}
\end{figure}

\textit{Reference frame and random-party distillation} - 
We now aim to address an intriguing question: if we do not distribute entanglement resources and instead perform measurements locally at each telescope, is there any advantage to pre-processing the stellar photons? To answer this question, we want to first point out that local measurements used for interferometry will consume some type of resource distributed from a central source, which is used to establish a shared reference frame. As outlined in Ref.~\cite{bartlett2007reference}, a state $\ket{\psi} = \alpha\ket{0} + \beta\ket{1}$ at telescope $X$ will be described at telescope $Y$ as $\hat{U}_\phi\ket{\psi} = \alpha\ket{0} + e^{i\phi}\beta\ket{1}$, with $\hat{U}_\phi = e^{i\phi\hat{n}}$, where $\phi$ represents the phase difference between Alice's and Bob's frames. Without a shared phase reference,  telescope $Y$ describes the state as $\int d\phi \hat{U}_\phi\ket{\psi}\bra{\psi}\hat{U}\phi^\dagger$, destroying coherence between photon-number bases. Superselection rule prevents local measurements on superpositions of different photon numbers without a shared phase reference.  A shared reference can be established by distributing $\ket{+}_X\ket{+}_Y$ to the two telescopes, where each $\ket{+}=(\ket{0}+\ket{1})/\sqrt{2}$ maintains the same relative phase between the vacuum and single-photon states. A coherent measurement can then be performed locally by incorporating these states into a local projection onto the single-photon sectors \cite{vsafranek2015quantum,bartlett2006degradation,bartlett2007reference}. 
The central server's task is to distribute $\ket{+}_X\ket{+}_Y$ to a pair of telescopes at each step, and server must decide which pair will receive the shared reference. We find that pre-processing with quantum randomness improves performance with the same amount of distributed resources.

%The goal then becomes for a central server to distribute $\ket{+}_X\ket{+}_{Y}$ to a pair of telescopes at each time step in the protocol.  Like before, the server must decide which pair of telescopes should receive the shared frame. We find that performance can again be enhanced by using quantum randomness from pre-processing, while consuming the same amount of distributed resources.

%In the discussion of previous sections, we assume we will implement the measurement assisted by entanglement distributed to distant telescopes as proposed in Ref.~\cite{gottesman2012longer}. One further interesting question is that if we do not distribute entanglement resource and only do measurement locally at each telescopes which consumes resources to build the shared reference frame, is there any advantage for doing the pre-processing for stellar photons. And we find that we can also enhance the performance using quantum randomness for the same amount of distributed resources.

%And we find that even if we want to implement local measurement, we can also save the resources distributed to the distant telescopes which are used to provide a shared reference frame between distant telescopes.

We first calculate the performance of estimating the coherence functions by local measurements with \textit{classical randomness} at the central server. In this case, the server flips a uniform coin and distributes a common reference frame to a randomly chosen pair of telescopes $XY$.  The parties can then locally project onto a coherent basis $\ket{\delta,\alpha}=(\ket{0}+ (-1)^{\alpha}e^{i\delta}\ket{1})/\sqrt{2}$, where $\alpha=0,1$, $\ket{0},\ket{1}$ are the vacuum and single photon states at the spatial mode corresponding to the given telescope.  The total measurement has POVM elements of the form
\begin{align}
E_{XY}&\!=\! \tfrac{1}{\binom{M}{2}}\op{\delta_X,\alpha_X}{\delta_X,\alpha_X}\!\otimes\op{\delta_Y,\alpha_Y}{\delta_Y,\alpha_Y}\!\otimes  \mathbb{I}_{\overline{XY}},\notag
\end{align} 
As derived in Sec.~\ref{SI:local_classical} of the Supplemental Material, the Fisher information for the estimation of each coherence function $g_{XY}$ has the block form
\begin{align}\label{classical_local_FI2}
F_{XY}^{cr}&=\frac{16\epsilon^2}{M^3(M-1)}\begin{pmatrix}\cos^2\delta_{XY}&\sin\delta_{XY}\cos\delta_{XY}\\\sin\delta_{XY}\cos\delta_{XY}&\sin^2\delta_{XY}\end{pmatrix},
\end{align}
%Note that to implement the measurement described here, we have to build a shared phase reference frame between telescopes $XY$ \cite{bartlett2007reference}. 
 Note that as pointed out in Ref.~\cite{tsang2011quantum}, when we want to do a local measurement, the performance is worse than the nonlocal measurement by a factor of $\epsilon$ due to the vacuum noise, which is clear by comparing the Fisher information for the local measurement in Eq.~\ref{classical_local_FI2} and nonlocal measurement in Eq.~\ref{classical_three}.

We now consider local measurements with \textit{quantum randomness} used to distribute a shared reference.  Like before, the stellar photon is pre-processed using the weak measurement in Eq.~\ref{eq:weak_measurement} before projecting onto $\ket{\delta,\alpha}_X=(\ket{0}+ (-1)^{\alpha}e^{i\delta}\ket{1})/\sqrt{2}$. After $D$ rounds of weak measurements, the corresponding POVM for telescopes $XY$ is given by
\begin{equation}\label{EMABlocal}
\begin{aligned}
&E_{XY}\!=\!\left[\tfrac{\beta_D}{4}\ket{00}\bra{00}+\tfrac{\gamma_D}{2}\ket{\psi}\bra{\psi}\right]_{XY}\!\!\!\!\!\otimes\op{0\cdots 0}{0\cdots 0}_{\overline{XY}}\\
\end{aligned}
\end{equation}
where $\ket{\psi}_{XY}=\left(\ket{01}+(-1)^{\alpha_X+\alpha_Y}e^{i(\delta_X-\delta_Y)}\ket{10}\right)/\sqrt{2}$, $\gamma_D,\beta_D$ depend on $D$ and $\tau_j$ as detailed in Sec.~\ref{SI:local_quantum} of the Supplemental Material.  The Fisher information of each coherence function $g_{XY}$ has the  form
\begin{align}\label{quantum_local_FI2}
    F_{XY}^{qr}&=\frac{M(M-1)}{4}\frac{\gamma_D^2}{\beta_D}F_{XY}^{cr},
\end{align}
%\begin{align}\label{quantum_local_FI2}
    %F_{XY}^{qr}&=\frac{4\epsilon^2}{M^2}\frac{\gamma_D^2}{\beta_D(1-\epsilon)}\begin{pmatrix}\cos^2\delta_{XY}&\sin\delta_{XY}\cos\delta_{XY}\\\sin\delta_{XY}\cos\delta_{XY}&\sin^2\delta_{XY}\end{pmatrix},
%\end{align}
in the limit $\beta_D(1-\epsilon)\gg \epsilon\gamma_D/M$ and $\epsilon\ll 1$.

\begin{thm}
%The ratio of the Fisher information, with the same amount of resources used for the reference frame in the case of quantum randomness compared to classical randomness, is given by
\begin{equation}
\frac{\Vert F^{qr}\Vert}{\Vert F^{cr}\Vert}=M(M-1)/4,
\end{equation}
up to the leading order when $\epsilon\ll 1$, where $F^{cr}$ and $ F^{qr}$ denote the Fisher Information matrix using classical randomness and quantum randomness \textit{per pair of projective measurement onto $\ket{\delta_X,{\alpha_X}}_X\otimes\ket{\delta_Y,{\alpha_Y}}_Y$.}
\end{thm}
The proof of this theorem is given in Sec.~\ref{SI:local_quantum} of the Supplemental Material. We further find that one round of weak measurement pre-processing is sufficient to improve the Fisher information by a factor of $M(M-1)/4$ using quantum randomness, if we compare Eq.~\ref{classical_local_FI2} and Eq.~\ref{quantum_local_FI2}. 
We explicitly calculate the Fisher information with the optimal $\tau_j$ for clarity in Sec.~\ref{SI:direct_check} of the Supplemental Material. Note that the weak measurement considered here does not require a shared reference frame between telescopes. %Furthermore, with the pre-processing step, we will project to a pair of telescopes and then project onto $\ket{\delta_{\alpha_X}}^X\otimes\ket{\delta_{\alpha_Y}}^Y$, which is the step consuming resources used as reference frame. In this sense, for measurement of each temporal mode of stellar light, 
Only the projection onto $\ket{\delta_X,\alpha_X}_X\otimes\ket{\delta_Y,\alpha_Y}_Y$ requires a shared frame, and thus the same amount of resources is consumed when using classical versus quantum randomness.

\textit{Conclusion} - %We have showed that using quantum randomness by pre-processing the stellar light using randomness distillation can enhance the performance of interferometric imaging with the same amount of resources distributed to distant telescopes. This shows that quantum randomness will potentially be useful when we want to increase the scale of astronomical interferometer to include more telescopes. 
We discussed two types of distributed resources for long-distant quantum telescopy: shared entanglement and a shared reference frame. %For both types, the pre-processing procedure can reduce the requirement amount of resources by a factor of $M/2$, where $M$ is the number of telescopes. 
In both cases, we have found that there is an advantage in pre-processing the stellar light before beginning the interferometric measurement between two telescopes.  The pre-processing considered here is implemented by a series of local weak quantum measurements, and we have analyzed the optimal choice of parameters for this process.  The results of Ref. \cite{Liu-2023a} suggest that our method offers the optimal way to pre-process each stellar photon for subsequent use in the GJC protocol. 
Our work offers a new approach to enhance distributed quantum sensing, addressing the often-overlooked challenges of resource distribution \cite{zhang2021distributed}.

Putting aside the experimental challenges, one could consider a central server that distributes multipartite resources or bipartite states to all telescopes simultaneously. How would this affect the Fisher information? For entanglement-based protocols, we explore two variants. The first sends multiple pairs of bipartite entangled states in each time step, running the GJC protocol in parallel across all telescopes. This parallel approach still yields a Fisher information per terrestrial photon that is $M/2$ times lower than the Fisher information of our pre-processing scheme, as it requires distributing $O(M)$ additional entangled states per step to achieve the same total Fisher information by capturing stellar photons received by all pairs of telescopes. Alternatively, the server could send a single multipartite entangled state (e.g., a W state) to all telescopes, as already considered in Ref. \cite{gottesman2012longer}.  While the Fisher information still roughly matches the Fisher information per terrestrial photon of our pre-processed scheme, the latter only requires distributing bipartite entanglement, regardless of how many telescopes are used. In the case of a shared reference frame, the total Fisher information  obtained by distributing a single shared frame to all telescopes matches that of our bipartite protocol. However, in the former scenario, the reference frame is phase locked across all telescopes, whereas in the latter, it exists only between a single pair of telescopes. More details can be found in Sec.~\ref{SI:sufficient_resource} of the Supplemental Material.

The practical implementation of pre-processing in our scheme presents significant experimental challenges. Weak measurement requires controlled two-qubit operations involving ancillary qubits. Moreover, quantum memory is essential because, after obtaining the measurement outcome, the telescopes must inform the server which pair should receive the resources prior to distribution. Therefore, quantum memory is needed to store the stellar states while awaiting resource allocation. Nevertheless, once such capabilities become available, we believe our protocol can largely enhance the performance of distributed quantum sensing and multi-telescope interferometry.

\textit{Acknowledgements} - We would like to thank Yujie Zhang for helpful discussion. This work was supported by the 
National Science Foundation Grant No. 1936321 and No. 2326803.

\bibliographystyle{h-physrev}
\bibliography{random_distillation}
\newpage

\appendix
\onecolumngrid

\section{Implementation of weak measurement}\label{implementation_POVM}
To implement the following measurement on a telescope, for instance, telescope A, 
\begin{align}\label{quantum_nonlocal_FI}
    M_0&=\begin{pmatrix}\sqrt{1-\tau}&0\\0&1\end{pmatrix}&M_1&=\begin{pmatrix}\sqrt{\tau}&0\\0&0\end{pmatrix}.
\end{align}
we introduce an ancillary qubit initialized in the state $\ket{0}_a$ and apply the following unitary operation, potentially mediated through a light-matter interaction between the stellar light and a matter qubit \cite{cirac1997quantum}, 
\begin{equation}
\ket{0}\bra{0}_A\otimes U_a+\ket{1}\bra{1}_A\otimes\mbb{I}_a,\quad U_a=\left[\begin{matrix}
\sqrt{1-\tau} & -\sqrt{\tau}\\
\sqrt{\tau} & \sqrt{1-\tau}
\end{matrix}\right]
\end{equation}
using telecopse A as the control qubit. Measuring the ancillary qubit $a$ in the computational basis $\ket{0}$ and $\ket{1}$ implements the measurement $M_0$ and $M_1$ on telescope A, respectively. 

To be more concrete, consider the case that we do weak measurement at each telescope for the stellar light received by three telescopes $\rho_{ABC}$.
If exactly two of the parties obtain outcome $M_0$ and the other party obtains outcome $M_1$, then the two parties getting $M_0$ will end up remaining coherently correlated while the third party is uncorrelated. The post-measurement state is given by $\rho_{ABC}=\rho_{AB}\otimes\ket{0}\bra{0}_C$, where
\begin{equation}\begin{aligned}
    \rho_{AB} &=  \frac{1}{P}\left(\beta_1(1 - \epsilon) |00 \rangle \langle 00|_{AB} + \gamma_1\frac{\epsilon}{3} \begin{pmatrix}
    1 & g_{AB} \\
    g_{AB}^* & 1 \\
    \end{pmatrix}\right),
\end{aligned}\end{equation}
and $\beta_1=(1-\tau)^2\tau$, $\gamma_1=(1-\tau)\tau$,  $P=(1-\epsilon)\beta_1+2\gamma_1\epsilon/3$ is the probability of getting this outcome and normalizes the state.
 When this occurs, the central server distributes the dual-rail qubit to the entangled parties and they perform the two-party GJC measurement.  On the other hand, if all three parties obtain outcome $M_0$,  the post-measurement state is given by
\begin{equation}\begin{aligned}
    \rho_{ABC} &=  \frac{1}{P}\left((1-\tau)^3(1 - \epsilon) |000 \rangle \langle 000| + (1-\tau)^2\epsilon\rho^{(1)}_{ABC} \right),
\end{aligned}\end{equation}
where $P=(1-\epsilon)(1-\tau)^3+\epsilon(1-\tau)^2$ is the probability of getting this outcome and normalizes the state.   A failure occurs if at anytime during the protocol at least two parties obtain an outcome $M_1$.

\section{Entanglement and randomness distillation}
\subsection{GJC protocol with classical randomness}\label{SI:POVM_gottesman_classical}
\subsubsection{Three telescopes case}
\label{SI:POVM_gottesman_classical_three}
%The goal of long baseline telescopy is to estimate the coherence functions $\{g_{AB}, g_{AC}, g_{BC}\}$. 

We will begin with the GJC protocol with classical randomness involving three telescopes. First, we will provide a more detailed definition of Fisher information.
In general, if $\mbf{c}:=(c_i)_{i=1}^l$ are some unknown classical parameters of a quantum system, then an exprimenter can extract information of a parameter using a positive operator-valued measure (POVM) $\{\Pi_x\}_x$, which delivers measurement outcome $x$ with probability some conditional distribution $P(x|\mbf{c})$.  An estimator $\hat{c}_i$ is then used to guess the value of $c_i$ using the outcome data $x$.  Letting  $\mathbb{E}\left[(\hat{c}_i-c_i)(\hat{c}_j-c_j)\right]$ denote the elements of the covariance matrix, the celebrated Cram\'{e}r Rao bound says that every unbiased estimator satisfies the positive semi-definite matrix inequality holds, $V-F^{-1}\geq 0$, where
\[F_{ij}=\sum_x\frac{1}{P(x|\mbf{c})}\frac{\partial P(x|\mbf{c})}{\partial c_i}\frac{\partial P(x|\mbf{c})}{\partial c_j}\]
are elements of the Fisher information matrix \cite{helstrom1976quantum}.  We are interested in computing the Fisher information in estimating each of the coherence functions $\{g_{AB}, g_{AC}, g_{BC}\}$ \textit{per dual-rail terrestrial photon generated}.  There are six real parameters being estimated, and we label them as $\mbf{c}:=(c_i)_{i=1}^6=(a_{AB},b_{AB},a_{AC},b_{AC},a_{BC},b_{BC})$, where $g_{XY}=a_{XY}+ib_{XY}$.

Using classical randomness, the server sends a dual-rail photon of the form $\frac{1}{\sqrt{2}}(\ket{01}+\ket{10})_{XY}$ to one pair of telescopes $XY$ chosen by flipping a uniform classical coin. We then interfere the terrestrial photon with the stellar photon at telescopes $XY$ to implement the GJC protocol. In total, the POVMs for the GJC protocol with classical randomness are
\begin{align}
   &E_{0,XY}= \frac{1}{3}\op{00}{00}_{XY}\otimes\mbb{I}_{\overline{XY}}, \quad E_{1,XY}= \frac{1}{6}\op{01}{01}_{XY}\otimes\mbb{I}^{\overline{XY}}, \quad
   E_{2,XY}= \frac{1}{6}\op{10}{10}^{XY}\otimes\mbb{I}^{\overline{XY}},\\
   & E_{3,XY}= \frac{1}{6}\op{\delta_+}{\delta_+}^{XY}\otimes\mbb{I}^{\overline{XY}}, \quad
     E_{4,XY}= \frac{1}{6}\op{\delta_-}{\delta_-}^{XY}\otimes\mbb{I}^{\overline{XY}},  
\end{align}
where $XY\in\{AB,AC,BC\}$,  $\overline{XY}$ is the complement of set $XY$, and $\ket{\delta_{\pm}}=\frac{1}{\sqrt{2}}(\ket{01}\pm e^{i\delta}\ket{10})$. 
In this work, for convenience, we will write the system labels $XY$ as subscripts for states and as superscripts for POVMs. Given parameter values $\mbf{c}$, the outcome probabilities are specified through Born's rule as
\begin{align}
    p(i,XY|\mbf{c})=\tr[E_{i,XY}\rho_{ABC}].
\end{align}
From the probability distribution, we can calculate Fisher information $F$ which decomposes into three $2\times 2$ blocks of the form for parameters $a_{XY},b_{XY}$
\begin{align}\label{classical_three2}
    F_{XY}^{cr}=\frac{\epsilon}{9\left[1-Re(g_{XY}e^{-i\delta})^2\right]}\begin{pmatrix}\cos^2\delta&\sin\delta\cos\delta\\\sin\delta\cos\delta&\sin^2\delta\end{pmatrix};
\end{align}
i.e. $F=F_{AB}\oplus F_{AC}\oplus F_{BC}$.

\subsubsection{$M$ telescopes case}

For the $M$ telescope case, the received stellar state has the form
\begin{equation}\begin{aligned}
\label{rhoM}
    &\rho_{A_1\cdots A_M} = (1 - \epsilon) |00\cdots 0 \rangle \langle 00\cdots 0| +\epsilon \rho^{(1)}+O(\epsilon^2),\\
    %&\rho^{(1)}_{XX}=1/M,\quad \rho^{(1)}_{XY}=g_{XY}/M,\quad g_{XY}=g_{YX}^*.\\
    &\rho^{(1)} = \frac{1}{M} \begin{pmatrix}
    1 & g_{A_1A_2} & g_{A_1A_3} & \cdots & g_{A_1A_M} \\
    g_{A_1A_2}^* & 1 & g_{A_2A_3}   & \cdots & g_{A_2A_M}\\
    g_{A_1A_3}^* & g_{A_2A_3}^* & 1   & \cdots & g_{A_3A_M}\\
    \cdots & \cdots & \cdots & \cdots & \cdots\\
    g_{A_1A_M}^* & g_{A_2A_M}^* & g_{A_3A_M}^*   & \cdots & 1
    \end{pmatrix},
\end{aligned}\end{equation}
\iffalse
\begin{equation}\label{rhoM}
    \rho_{A_1\cdots A_M} = (1 - \epsilon) |00\cdots 0 \rangle \langle 00\cdots 0| +\epsilon \rho^{(1)}+O(\epsilon^2),\quad \rho^{(1)}_{ii}=1/M,\quad \rho^{(1)}_{i<j}=g_{ij}/M,\quad \rho^{(1)}_{i>j}=g_{ij}^*/M.
\end{equation}

This time the $AB$ reduced density matrix has the form
\begin{align}
    \rho_{AB}&=\left(1-\frac{2}{M}\epsilon\right)\op{00}{00}+\frac{2}{M}\epsilon\rho_{AB}^{(1)}+O(\epsilon^2)
\end{align}
\fi
where the coherence term is given by $g_{XY}=\int dx I(x)e^{ik_{XY}x}=a_{XY}+ib_{XY}$, where $I(x)$ is the intensity distribution of the source and $XY\in\{A_1A_2,A_1A_3,\cdots,A_{M-1}A_M\}$.  The parameter $k_{XY}$ is determined by the baseline between telescope $X$ and $Y$, the wavelength of the light, and the distance between the image and source planes.   Again, using classical randomness, the server sends a dual-rail photon of the form $\frac{1}{\sqrt{2}}(\ket{01}+\ket{10})_{XY}$ to one pair of telescopes $XY$ chosen by flipping a uniform classical coin. The relevant POVMs (i.e. those whose outcomes affect the Fisher Information) are given by
\begin{align}
E_{XY}&= \frac{1}{2\binom{M}{2}}\op{\delta_+}{\delta_+}_{XY}\otimes\mbb{I}_{\overline{XY}},\quad E_{XY}= \frac{1}{2\binom{M}{2}}\op{\delta_-}{\delta_-}_{XY}\otimes\mbb{I}_{\overline{XY}}.
\end{align}
The Fisher Information matrix \textit{per terrestrial photon} using classical randomness has blocks of the form
\begin{align}\label{classical_nonlocal_FI}
    F_{XY}^{cr}&=\frac{\epsilon}{M\binom{M}{2}\left[1-Re(g_{XY}e^{-i\delta})^2\right]}\begin{pmatrix}\cos^2\delta&\sin\delta\cos\delta\\\sin\delta\cos\delta&\sin^\delta\end{pmatrix},
\end{align}

\subsection{GJC protocol with quantum randomness}\label{SI:POVM_gottesman_quantum}

\subsubsection{Three telescopes case}

We now consider the GJC protocol with quantum randomness for three telescopes case.
To compute the Fisher Information matrix, it will be helpful to explicitly write down the POVMs.  Suppose the parties agree in advance to perform $D$ total rounds of measurement.  We let $\tau_r$ denote the strength of local measurement in the $r^{th}$ round of measurement.  Then the relevant POVMs (i.e. those whose outcomes affect the Fisher Information) are given by
\begin{align}
     &E_{XY}= \frac{\gamma_D}{2}\op{\delta_\pm}{\delta_\pm}_{XY}\otimes\op{0}{0}_{{\overline{XY}}}, 
\end{align}
\iffalse
\begin{align}
     E_3^{AB}&= \frac{\gamma_D}{2}\op{\delta_+}{\delta_+}^{AB}\otimes\op{0}{0}^C,  &E_3^{AC}&= \frac{\gamma_D}{2}\op{\delta_+}{\delta_+}^{AC}\otimes\op{0}{0}^B,&E_3^{BC}&= \frac{\gamma_D}{2}\op{\delta_+}{\delta_+}^{BC}\otimes\op{0}{0}^A\notag\\
     E_4^{AB}&= \frac{\gamma_D}{2}\op{\delta_-}{\delta_-}^{AB}\otimes\op{0}{0}^C,  &E_4^{AC}&= \frac{\gamma_D}{2}\op{\delta_-}{\delta_-}^{AC}\otimes\op{0}{0}^B,&E_4^{BC}&= \frac{\gamma_D}{2}\op{\delta_-}{\delta_-}^{BC}\otimes\op{0}{0}^A,
\end{align}
\fi
where 
\begin{equation}
    \gamma_D=\sum_{r=1}^D(1-\tau_r)\tau_r\left(\prod_{k=1}^{r-1}(1-\tau_{k})^2\right).
\end{equation}
For instance, POVM element $E^{AB}$ corresponds to the event that Charlie obtains outcome $M_1$ in one of the $D$ rounds of random distillation (while Alice and Bob both measure $M_0$), and then Alice and Bob measure $\frac{1}{2}\op{\delta_\pm}{\delta_\pm}_{XY}$ in the bipartite GJC protocol.  With this POVM, we can proceed to calculate the Fisher information
\begin{align}\label{quantum_three2}
    F_{XY}^{qr}=\frac{\gamma_D\epsilon}{3\left[1-Re(g_{XY}e^{-i\delta})^2\right]}\begin{pmatrix}\cos^2\delta&\sin\delta\cos\delta\\\sin\delta\cos\delta&\sin^2\delta\end{pmatrix}.
\end{align}
We want to find the largest $\gamma_D$ which gives the optimal performance.

Let's compute the largest $\gamma_D$ by optimizing $\tau_r$ in more detail.  For any specified pair of parties $XY$, the POVM element that uncouples all the other parties has the form $\gamma_D\mbb{I}_{XY}\otimes\op{0}{0}_{\overline{XY}}$, where
\begin{align}
    \gamma_D&=\omega_{D-1}+\frac{1}{\binom{3}{2}}\prod_{k=1}^{D-1}(1-\tau_{k})^2,
\end{align}
where 
\begin{equation}
    \omega_{d}=\sum_{r=1}^{d}\prod_{k=1}^{r-1}(1-\tau_{k})^2(1-\tau_r)\tau_r.
\end{equation}
We then write
\begin{align}
    \gamma_D&=\omega_{D-1}+\frac{1}{\binom{3}{2}}\prod_{k=1}^{D-1}(1-\tau_{k})^2\notag\\
    &=\omega_{D-2}+\frac{1}{3}\prod_{k=1}^{D-2}(1-\tau_k)^2(3(1-\tau_{D-1})\tau_{D-1}+(1-\tau_{D-1})^2)\notag\\
    &=\omega_{D-2}+\frac{1}{3}\prod_{k=1}^{D-2}(1-\tau_k)^2(1+\tau_{D-1}-2\tau_{D-1}^2)
\end{align}
Note that $\omega_{D-1}$ does not depend on $\tau_{D-1}$.  Hence, we can easily compute the optimal $\tau_{D-1}$ by taking
\begin{align}
   0= \frac{\partial\gamma_D}{\partial \tau_{D-1}}&=\frac{1}{3}\prod_{k=1}^{D-2}(1-\tau_{k})^2\left[1-4\tau_{D-1}\right]\quad\Rightarrow\tau_{D-1}=\frac{1}{4}.
  \end{align}
  Therefore, our formula for $\gamma_D$ becomes
  \begin{align}
      \gamma_{D}&=\omega_{D-2}+\frac{1}{3}\prod_{k=1}^{D-2}(1-\tau_k)^2\frac{9}{8}\notag\\
      %&=\omega_{D-3}+\frac{1}{3}\prod_{k=1}^{D-3}(1-\tau_k)^2(3(1-\tau_{D-2})\tau_{D-2}+\frac{9}{8}(1-\tau_{D-2})^2)\notag\\
      &=\omega_{D-3}+\frac{1}{3}\prod_{k=1}^{D-3}(1-\tau_k)^2(\frac{3}{8}(3+2\tau_{D-2}-5\tau_{D-2}^2)).
  \end{align}
  This is maximized with the choice $\tau_{D-2}=\frac{1}{5}$, which yields
  \begin{align}
      \gamma_D=\omega_{D-3}+\frac{1}{3}\prod_{k=1}^{D-3}(1-\tau_k)^2\frac{6}{5}.
  \end{align}
  Following the pattern, we have
  \begin{align}
      \gamma_D&=\omega_{D-2}+\frac{1}{3}\prod_{k=1}^{D-2}(1-\tau_k)^2\frac{9}{8}\notag\\
      \gamma_D&=\omega_{D-3}+\frac{1}{3}\prod_{k=1}^{D-3}(1-\tau_k)^2\frac{6}{5}\notag\\
      \gamma_D&=\omega_{D-4}+\frac{1}{3}\prod_{k=1}^{D-4}(1-\tau_k)^2\frac{5}{4}\notag\\
      \gamma_D&=\omega_{D-5}+\frac{1}{3}\prod_{k=1}^{D-5}(1-\tau_k)^2\frac{9}{7}\notag\\
      \cdots \quad\gamma_D&=\omega_{D-k}+\prod_{k=1}^{D-k}(1-\tau_k)^2\frac{k+1}{2(k+2)}.
  \end{align}
  Taking $k=D$ and using the fact that $\omega_{0}=0$, we have the formula
  \begin{align}
      \gamma_D=\frac{D+1}{2(D+2)}.
  \end{align}

\subsubsection{$M$ telescopes case}\label{SI:GJC M}
For the $M$ telescopes case, the relevant POVM of the GJC protocol with quantum randomness has a similar form
\begin{align}\label{eq:POVM_Gottesman_quantum_M}
E_{XY}&= \frac{\gamma_D}{2}\op{\delta_\pm}{\delta_\pm}_{XY}\otimes\op{00\cdots0}{00\cdots0}_{\overline{XY}},\\
%E_{XY}&= \frac{\gamma_D}{2}\op{\delta_-}{\delta_-}_{XY}\otimes\op{00\cdots0}{00\cdots0}_{\overline{XY}},
\end{align}
We will show below that with the optimal choice of $\tau_r$,  we have $\gamma_D=1/(M-1)$. Since we only use one terrestrial photon per temporal mode, the Fisher information \textit{per terrestrial photon} using quantum randomness is
\begin{align}\label{eq:FI_gottesman_quantum_M}
    F_{XY}^{qr}=\frac{\gamma_D\epsilon}{M\left[1-Re(g_{XY}e^{-i\delta})^2\right]}\begin{pmatrix}\cos^2\delta&\sin\delta\cos\delta\\\sin\delta\cos\delta&\sin^2\delta\end{pmatrix}
\end{align}
To find $\gamma_D$, we need to find the POVMs of performing $D$ total rounds of measurement. Assume the strength of the local measurement at each telescope is the same in each round. Let $\tau_r$ denote the strength of local measurement in the $r^{\text{th}}$ round of measurement. The POVM of projecting onto telescopes A and B at $r$th measurement for $M$ telescopes is given as
\begin{equation}
\label{Eq:M-r}
E^{}_{r,M,\text{AB}}=\left[\prod_{j=1}^r E_0(\tau_j)\right]\otimes\left[\prod_{j=1}^r E_0(\tau_j)\right]\otimes e^{M-2}
\end{equation}
where we have $E_0$ at telescopes A and B in all $r$ rounds of measurement, $e^{M-2}$ includes the operators for the other $M-2$ telescopes. Here, $e^{M-2}$ must have at least one telescope that gets the $E_1$ outcome for the first time at the $r$th measurement, %, i.e. $E_0\otimes E_0\otimes \cdots \otimes E_0\otimes E_1$. 
and all telescopes in $e^{M-2}$ must have at least one $E_1$ outcome.  For ease of calculation, at each round, all telescopes will consistently project onto $E_{0,1}$, regardless of the prior measurement outcomes. We will find the explicit form of $e^{M-2}$ inductively below.

 Now, we add another telescope and find $E^{}_{r,M+1,\text{AB}}$. Importantly,  $E_{r,M+1,\text{AB}}$ refers to the measurement which successfully projects onto telescopes $A,B$ at the $r$th round of measurement. All $M-1$ telescopes other than telescopes A and B  (the 3rd to $M+1$th telescopes)  have at least one $E_1$ among themselves, and at least one of them gets $E_1$ for the first time at the $r$th measurement. When adding a new telecopes to $M$ telecopes to find $E_{r,M+1,\text{AB}}$, we will classify into two cases depending on the position of telescopes which measures $E_1$ for the first time at $r$th round of measurement. Specifically, the telescope that first measures $E_1$ at the $r$th round could either be the newly added $(M+1)$th telescope or one of the existing $M-2$ telescopes (i.e., the $3$rd through $M$th telescopes):

First case is that we make sure  at least one of the telescope in the $3$rd to $M$th telescopes measures $E_1$ for the first time at $r$th round of measurement. And the added $(M+1)$th telescope only needs to have at least one $E_1$ outcome, i.e. $I-\prod_{j=1}^r E_0(\tau_j)$. The POVMs for the $M-1$ telescopes other than telescopes A and B is $e^{M-2}\otimes\left(I-\prod_{j=1}^r E_0(\tau_j)\right)$ in this case. 

Second case is that the $3$rd to $M$th telescopes get $E_1$ earlier than the $r$th measurement described by $\left(I-\prod_{j=1}^r E_0(\tau_j)\right)\otimes\left(I-\prod_{j=1}^r E_0(\tau_j)\right)\otimes\cdots\otimes\left(I-\prod_{j=1}^r E_0(\tau_j)\right)-e^{M-2}$. And the added $(M+1)$th telescope gets $E_1$ at $r$th measurement for the first time described by $\prod_{j=1}^{r-1} E_0(\tau_j)E_1(\tau_r)$.

\begin{figure}[!tbh]
\begin{center}
\includegraphics[width=0.4\columnwidth]{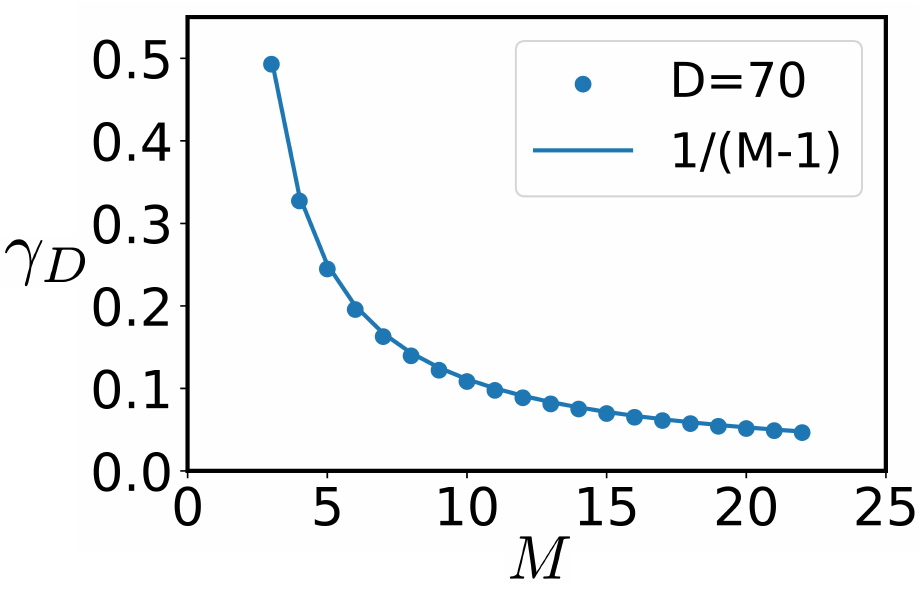}
\caption{$\gamma_D$ as a function of the number of telescopes $M$ with the numerically optimized  $\tau_j$. The dot indicates the numerical calculation of optimal $\gamma_D$ when $D=70$. The solid line shows the fitted formula  $\gamma_D=1/(M-1)$.} 
\label{gamma_D_M}
\end{center}
\end{figure}

\begin{figure}[!tbh]
\begin{center}
\includegraphics[width=0.4\columnwidth]{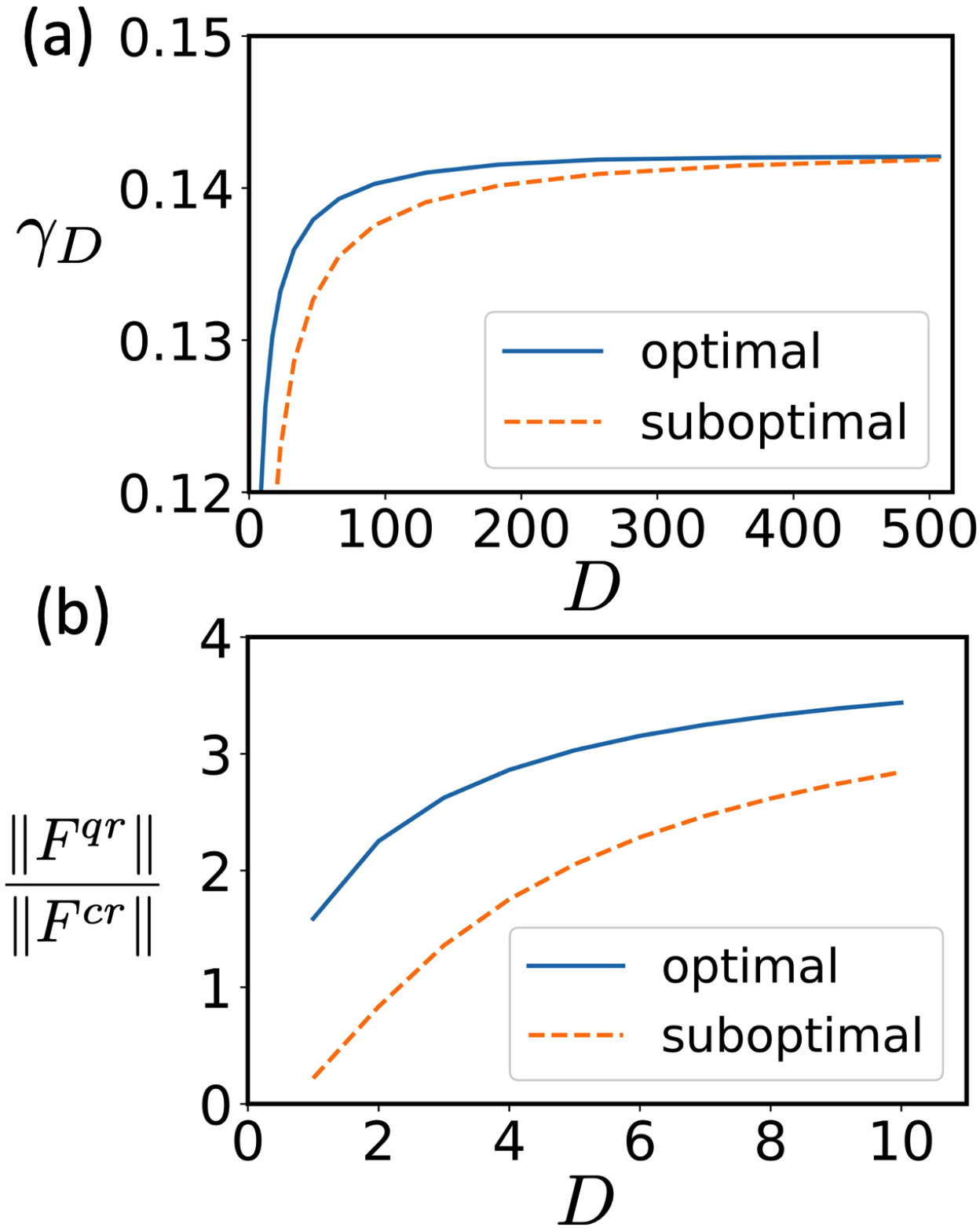}
\caption{(a) $\gamma_D$ as a function of the number of measurement rounds $D$. (b) Ratio between Fisher information per terrestrial photon with quantum and classical randomness $\frac{\Vert F^{qr}\Vert}{\Vert F^{cr}\Vert}$ as a fuction of the number of measurement rounds $D$. Both are calculated with the numerically optimized  $\tau_j$ (solid line) and a suboptimal choice of $\tau_j=\frac{1}{2+D-j}$ (dash line). Number of telescopes $M=8$.  } 
\label{suboptimal}
\end{center}
\end{figure}

\begin{figure}[!tbh]
\begin{center}
\includegraphics[width=0.45\columnwidth]{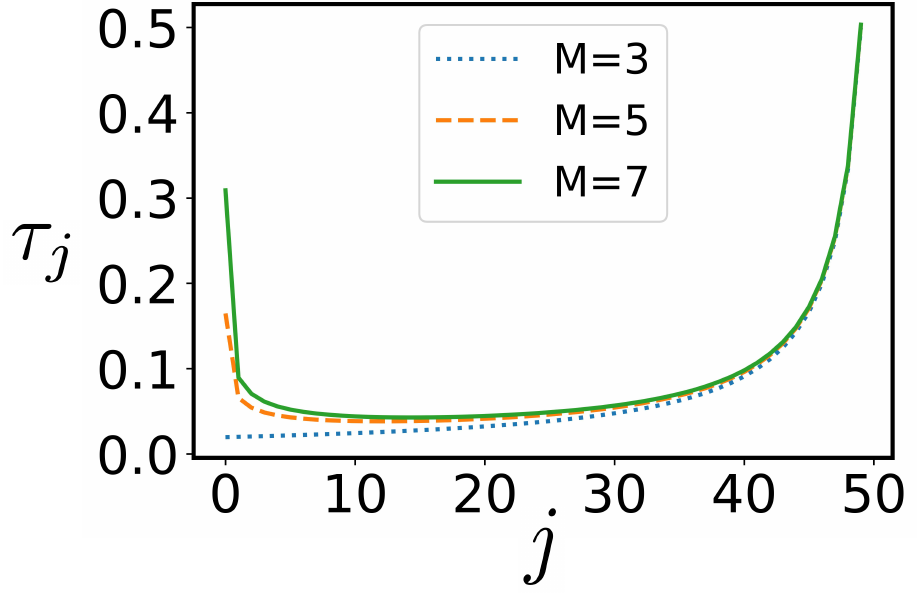}
\caption{Optimized $\tau_j$ for different number of telescopes $M=3,5,7$ when the number of measurement rounds $D=50$. } 
\label{choice_tau}
\end{center}
\end{figure}

We have thus derived the form of $E^{}_{r,M+1,\text{AB}}$,
\begin{equation}\label{ErM}
\begin{aligned}
E^{}_{r,M+1,\text{AB}}&=\left[\prod_{j=1}^r E_0(\tau_j)\right]\otimes\left[\prod_{j=1}^r E_0(\tau_j)\right]\otimes \Bigg[e^{M-2}\otimes\left(I-\prod_{j=1}^r E_0(\tau_j)\right)\\
&+ \left(\left(I-\prod_{j=1}^r E_0(\tau_j)\right)\otimes\left(I-\prod_{j=1}^r E_0(\tau_j)\right)\otimes\cdots\otimes\left(I-\prod_{j=1}^r E_0(\tau_j)\right)-e^{M-2}\right)\otimes\left(\prod_{j=1}^{r-1} E_0(\tau_j)E_1(\tau_r)\right)\Bigg] 
\end{aligned}
\end{equation}
$\gamma_D$ can be calculated from this relation
\begin{equation}
\begin{aligned}
&\gamma_D=\sum_{r=1}^D\gamma_{D,r},\quad x_{r\geq 1}=\prod_{j=1}^r(1-\tau_j),\quad x_0=1\\
&\gamma_{D,r}=\bra{1000\cdots}E^{}_{r,M,\text{AB}}\ket{1000\cdots}=\bra{0100\cdots}E^{}_{r,M,\text{AB}}\ket{0100\cdots}\\
&=\left[1-\prod_{j=1}^{r-1}(1-\tau_j)\right]\bra{1000\cdots}E_{r,M=3,\text{AB}}\ket{1000\cdots}\\
&\quad\quad+\left(\prod_{j=1}^{r-1}(1-\tau_j)^2\right)\tau_r(1-\tau_r)\sum_{n=0}^{M-4}\left[1-\prod_{j=1}^{r-1}(1-\tau_j)\right]^n\left[1-\prod_{j=1}^r(1-\tau_j)\right]^{M-3-n}\\
%&=\prod_{j=1}^{r-1}(1-\tau_j)^2(1-\tau_r)\tau_r\left\{\left[1-\prod_{j=1}^{r-1}(1-\tau_j)\right]^{M-3}+\left[1-\prod_{j=1}^{r}(1-\tau_j)\right]\frac{\left[1-\prod_{j=1}^{r}(1-\tau_j)\right]^{M-3}-\left[1-\prod_{j=1}^{r-1}(1-\tau_j)\right]^{M-3}}{\prod_{j=1}^{r-1}(1-\tau_j)-\prod_{j=1}^{r}(1-\tau_j)}\right\}\\
&=x_r\left[(1-x_r)^{M-2}-(1-x_{r-1})^{M-2}\right],\\
\end{aligned}
\end{equation}
It is clear that $\gamma_D$ is a function of the number of telescopes $M$, the number of measurement rounds $D$, and the choice of strength of measurement $\tau_{j=1,2,\cdots, D}$ in each round. We want to find the optimization over $\tau_{j=1,2,\cdots, D}$ such that $\gamma_D$ achieves its maximal value. Unfortunately, we find the analytical calculation is hard. The numerical calculation for $\gamma_D$ with numerically optimized $\tau_r$ in each round is shown in Fig.~\ref{gamma_D_M}.  It turns out that $\gamma_D=\frac{1}{M-1}$ fits the calculation results very well.
We have included a plot for $\gamma_D$ and ${\Vert F^{qr}\Vert}/{\Vert F^{cr}\Vert}$ as a fuction of the number of measurement rounds $D$ in Fig.~\ref{suboptimal}. The numerically optimized $\tau_j$ is shown in Fig.~\ref{choice_tau} for different number of telescopes $M$. For the case of $M=3$, the numerical results of $\tau_j$ matches exactly the analytical calculation above and $\tau_j$ should be chosen to be a small values in the beginning and increases until the last round of measurement. But when the number of telescopes $M\geq 4$, optimal choice of $\tau_j$ is large in the beginning, decreases in the middle, and increases again in the end. Intuitively, the optimal strategy first quickly collapses the states to a small number of telescopes and then slowly collapses to the last two telescopes.

\section{Reference frame and randomness distillation}
\subsection{Local scheme with classical randomness}\label{SI:local_classical}

We now consider the local scheme with classical randomness in the $M$ telescopes case.
The POVM element whose outcomes affect the Fisher information is
\begin{align}
E(\alpha_X,\alpha_Y)_{XY}&= \frac{1}{\binom{M}{2}}\op{\delta_X,\alpha_X}{\delta_X,\alpha_X}_{X}\otimes\op{\delta_Y,\alpha_Y}{\delta_Y,\alpha_Y}_{Y}\otimes  I_{\overline{XY}},
\end{align}
For the stellar state described in Eq.~\ref{rhoM}, the probability of getting outcome described by $\alpha_X,\alpha_Y$, 
\begin{equation}
P(\alpha_X,\alpha_Y)=\frac{1}{M(M-1)}\left[1+\frac{2\epsilon}{M}|g_{XY}|\cos(\theta+\delta_{XY})(-1)^{\alpha_X+\alpha_Y}\right]
\end{equation}
where $\delta_{XY}=\delta_X-\delta_Y$, $M$ is the number of telescopes.  The Fisher information for the estimation of each coherence function $g_{XY}$ has the block form
\begin{align}\label{classical_local_FI}
    F_{XY}^{cr}&=\sum_{\alpha_X,\alpha_Y}\frac{(2\epsilon/M^2(M-1))^2}{P(\alpha_X,\alpha_Y)}\begin{pmatrix}\cos^2\delta_{XY}&\sin\delta_{XY}\cos\delta_{XY}\\\sin\delta_{XY}\cos\delta_{XY}&\sin^2\delta_{XY}\end{pmatrix}\\
    &=\frac{16\epsilon^2}{M^3(M-1)}\begin{pmatrix}\cos^2\delta_{XY}&\sin\delta_{XY}\cos\delta_{XY}\\\sin\delta_{XY}\cos\delta_{XY}&\sin^2\delta_{XY}\end{pmatrix},
\end{align}
where we ignore the subleading order terms of $\epsilon$ in $P(\alpha_X,\alpha_Y)$ from the first row to the second row, and $M$ is the number of telescopes.

\subsection{Local scheme with quantum randomness}\label{SI:local_quantum}

For the local scheme with quantum randomness in the $M$ telescopes case,
to calculate the Fisher information of a local scheme after pre-processing using a sequence of weak measurement, we need to include the vacuum terms in the POVM $E^{}_{r,M,\text{AB}}$ given in Eq.~\ref{ErM},
\begin{equation}
\begin{aligned}
\beta_{D,r}:=\bra{0000\cdots}E^{}_{r,M,\text{AB}}\ket{0000\cdots}=\prod_{j=1}^r(1-\tau_j)\bra{1000\cdots}E^{}_{r,M,\text{AB}}\ket{1000\cdots}=\gamma_{D,r}x_r,\quad \beta_D=\sum_{r=1}^D\beta_{D,r}
\end{aligned}
\end{equation}
\begin{equation}\label{EMABlocal}
\begin{aligned}
&E_{M,\text{AB}}=\beta_D\ket{0000\cdots}\bra{0000\cdots}+\gamma_D\left(\ket{1000\cdots}\bra{1000\cdots}+\ket{0100\cdots}\bra{0100\cdots}\right)+\cdots
\end{aligned}
\end{equation}
If we first do the weak measurement using $E_{M,{XY}}$ and then project onto $\ket{\delta_{X},\alpha_X}\otimes\ket{\delta_{Y},\alpha_Y}$, $\alpha_{X,Y}=0,1$, the probability distribution of getting each outcome is given by
\begin{equation}
\begin{aligned}
&P(++,--)=\frac{\beta_D}{2}(1-\epsilon)+\frac{\epsilon\gamma_D}{M}(1+\text{Re}g_{XY}\cos\delta_{XY}+\text{Im}g_{XY}\sin\delta_{XY})\\
&P(+-,-+)=\frac{\beta_D}{2}(1-\epsilon)+\frac{\epsilon\gamma_D}{M}(1-\text{Re}g_{XY}\cos\delta_{XY}-\text{Im}g_{XY}\sin\delta_{XY})
\end{aligned}
\end{equation}
And the Fisher information of each coherence function $g_{XY}$ has the block form
\begin{align}\label{quantum_local_FI}
    F_{XY}^{qr}
    &=\frac{\epsilon^2\gamma_D^2}{M^2}\frac{\beta_D(1-\epsilon)+\frac{2\epsilon}{M}\gamma_D}{(\frac{\beta_D}{2}(1-\epsilon)+\frac{\epsilon}{M}\gamma_D)^2-\frac{\epsilon^2\gamma_D^2}{M^2}(\text{Re}g_{XY}\cos\delta_{XY}+\text{Im}g_{XY}\sin\delta_{XY})^2}\begin{pmatrix}\cos^2\delta_{XY}&\sin\delta_{XY}\cos\delta_{XY}\\\sin\delta_{XY}\cos\delta_{XY}&\sin^\delta_{XY}\end{pmatrix}\\
    &\leq\frac{4\epsilon^2}{M^2}\frac{\gamma_D^2}{\beta_D(1-\epsilon)}\frac{\beta_D(1-\epsilon)+\frac{2\epsilon}{M}\gamma_D}{\beta_D(1-\epsilon)+\frac{4\epsilon}{M}\gamma_D}\begin{pmatrix}\cos^2\delta_{XY}&\sin\delta_{XY}\cos\delta_{XY}\\\sin\delta_{XY}\cos\delta_{XY}&\sin^2\delta_{XY}\end{pmatrix}\\
    &\leq\frac{4\epsilon^2}{M^2}\frac{\gamma_D^2}{\beta_D(1-\epsilon)}\begin{pmatrix}\cos^2\delta_{XY}&\sin\delta_{XY}\cos\delta_{XY}\\\sin\delta_{XY}\cos\delta_{XY}&\sin^2\delta_{XY}\end{pmatrix},
\end{align}
The two inequalities are  saturated when $\beta_D(1-\epsilon)\gg \epsilon\gamma_D/M$. And we will consider our scheme in weak thermal source limit $\epsilon\ll1$, which is common in optical wavelength.

Let us now consider the choice of $\tau_r$ which optimize $\gamma_D^2/\beta_D$. We want to first prove  that $\gamma_D^2/\beta_D\leq 1$ and then show this inequality is saturable. To prove the inequality, we notice  $0\leq\gamma_{D,r}\leq x_r\leq 1$ which is clear since $0\leq\tau_j\leq1$. Furthermore, we can observe that
\begin{equation}
\sum_{r=1}^D\gamma_{D,r}/x_r=(1-x_D)^{M-2}\leq 1
\end{equation}
We have
\begin{equation}
\beta_D\geq \beta_D\left(\sum_{r=1}^D\gamma_{D,r}/x_r\right)=\left(\sum_{r=1}^D\gamma_{D,r}x_r\right)\left(\sum_{r=1}^D\gamma_{D,r}/x_r\right)\geq\left(\sum_{r=1}^D\gamma_{D,r}\right)^2=\gamma_D^2
\end{equation}
where the second inequality follows from the Cauchy–Schwarz inequality. The first inequality is saturated when $x_D\rightarrow 0$. The second inequality is saturated when $x_r$ is the same for  $\forall r$. So, we want to choose $\tau_r$ such that $x_r\rightarrow0$ for  $\forall r$. And this can be easily satisfied if we use $D=1$ and $\tau_1\rightarrow 1$. We want to emphasize that we cannot choose $\tau_1=1$, instead we must make sure $1-\tau_1\gg \epsilon/M$, since we also want to saturate the inequality in Eq.~\ref{quantum_local_FI} which requires $\beta_{D=1}(1-\epsilon)\gg\epsilon\gamma_{D=1}/M$. Intuitively, this means we want to do one round weak measurement which is very close to a projective measurement. But the measurement cannot be a hard projective measurement where $\tau_1=1$ because in this case we directly project onto the vacuum or single photons states at each telescope and destroy the coherence between telescopes. And when $D=1$, $\tau_1=1$, we have
$\gamma_{D=1}=\beta_{D=1}=0$, this is because in this case $E_{M,{AB}}=\ket{1100\cdots0}\bra{1100\cdots0}$, which does not include the  basis states considered in Eq.~\ref{EMABlocal}.

\subsection{Direct calculation of the local scheme using quantum randomness}\label{SI:direct_check}

For a single round of weak measurement, the weak measurement can be described as follows (assuming, without loss of generality, that the system collapses to telescopes $A$ or $B$):
\begin{equation}\begin{aligned}
&M=M_0\otimes M_0 \otimes M_1\otimes \cdots\otimes M_1,\\
&M^\dagger M=\tau^{M-2}[(1-\tau)^2\ket{00\cdots 0}\bra{00\cdots0}+(1-\tau)\ket{10\cdots 0}\bra{10\cdots0}+(1-\tau)\ket{01\cdots 0}\bra{01\cdots0}+\ket{11\cdots 0}\bra{11\cdots0}].
\end{aligned}\end{equation}
The post-measurement state corresponding to this weak measurement outcome, with probability $\tr(M\rho M^\dagger)$, can be described as:
\begin{equation}\begin{aligned}
\rho'&=\frac{M\rho M^\dagger}{\tr(M\rho M^\dagger)}\\
&=\frac{1}{\tr(M\rho M^\dagger)}[(1-\tau)^2\tau^{M-2}(1-\epsilon)\ket{00\cdots 0}\bra{00\cdots0}+(1-\tau)\tau^{M-2}\frac{\epsilon}{M}\begin{pmatrix}1 & g_{AB}\\ g_{AB}^* &1\end{pmatrix}],
\end{aligned}\end{equation}
where the second term is expressed in the basis $\ket{100\cdots0}, \ket{010\cdots0}$ for telescopes $A$ and $B$. Note that we have neglected the $O(\epsilon^2)$ terms, which scale as $\tau^{M-2}\epsilon^2/M^2$, since their contribution is significantly smaller than the $O(\epsilon)$ terms in the limit where $1-\tau \gg \epsilon/M$, as considered here. We then project onto $\ket{\delta_{A},\alpha_A}\otimes\ket{\delta_{B},\alpha_B}$, $\alpha_{A,B}=0,1$,, and the probability distribution is given by
\begin{equation}\label{eq:P}
\begin{aligned}
&P(++,--)=\frac{1}{2}(1-\tau)\tau^{M-2}\left[(1-\epsilon)(1-\tau)+\frac{2\epsilon}{M}(1+\text{Re}g_{AB}\cos\delta_{AB}+\text{Im}g_{AB}\sin\delta_{AB})\right],\\
&P(+-,-+)=\frac{1}{2}(1-\tau)\tau^{M-2}\left[(1-\epsilon)(1-\tau)+\frac{2\epsilon}{M}(1-\text{Re}g_{AB}\cos\delta_{AB}-\text{Im}g_{AB}\sin\delta_{AB})\right].
\end{aligned}
\end{equation}
The Fisher information for local scheme with quantum randomness is given by
\begin{equation}
\begin{aligned}\label{eq:direct_FI}
    F_{AB}^{qr}&=\left((1-\tau)\tau^{M-2}\frac{\epsilon}{M}\right)^2\begin{pmatrix}\cos^2\delta_{AB}&\sin\delta_{AB}\cos\delta_{AB}\\\sin\delta_{AB}\cos\delta_{AB}&\sin^2\delta_{AB}\end{pmatrix}\\
    &\quad \times \frac{(1-\tau)\tau^{M-2}\left[(1-\epsilon)(1-\tau)+\frac{2\epsilon}{M}\right]}{(\frac{1}{2}(1-\tau)\tau^{M-2})^2[((1-\epsilon)(1-\tau)+\frac{2\epsilon}{M})^2-(\frac{2\epsilon}{M})^2(\text{Re}g_{XY}\cos\delta_{XY}+\text{Im}g_{XY}\sin\delta_{XY})^2]}\\
    &=\frac{4\epsilon^2}{M^2}\frac{(1-\tau)\tau^{M-2}(1-\epsilon)(1-\tau)}{(1-\epsilon)^2(1-\tau)^2}\begin{pmatrix}\cos^2\delta_{AB}&\sin\delta_{AB}\cos\delta_{AB}\\\sin\delta_{AB}\cos\delta_{AB}&\sin^2\delta_{AB}\end{pmatrix}\\
    &=
    \frac{4\epsilon^2}{M^2}\begin{pmatrix}\cos^2\delta_{AB}&\sin\delta_{AB}\cos\delta_{AB}\\\sin\delta_{AB}\cos\delta_{AB}&\sin^2\delta_{AB}\end{pmatrix},
\end{aligned}    
\end{equation}
where we have used the limit $\epsilon\rightarrow0$, $\tau\rightarrow1$, $1-\tau\gg\epsilon/M$. When compared to the case of using classical randomness in Eq.~\ref{classical_local_FI}
\begin{align}\label{classical_local_FI3}
F_{XY}^{cr}&=\frac{16\epsilon^2}{M^3(M-1)}\begin{pmatrix}\cos^2\delta_{XY}&\sin\delta_{XY}\cos\delta_{XY}\\\sin\delta_{XY}\cos\delta_{XY}&\sin^2\delta_{XY}\end{pmatrix},
\end{align}
the quantum randomness leads to an improvement by a factor of $M(M-1)/4$, as anticipated. %In this way, we have directly demonstrated the advantage.

Note that compared to the ratio between the GJC protocol with quantum and classical randomness per terrestrial photon is $M/2$, the ratio per pair of projective measurement considered here has a larger factor $M(M-1)/4$. Intuitively, this is because the weak measurement described in Eq.~\ref{EMABlocal} can suppress the vacuum terms of stellar light, which is an important source of noise for the local projective measurement \cite{tsang2011quantum}.

%We also want to emphasize that in Eq.~\ref{eq:P}, the term from vacuum term is suppressed by a factor of $1-\tau$ when $\tau\rightarrow1$, which introduce an extra factor of $1-\tau$ in the denominator of Eq.~\ref{eq:direct_FI}. Since we need to take derivative in the numerator of Eq.~\ref{eq:direct_FI}, this extra factor does not appear in the numerator, which makes the Fisher information independent of $\tau$ up to the leading order. Compared to the Fisher information for the Gottesman protocol with quantum randomness in Eq.~\ref{eq:FI_gottesman_quantum_M}, the Fisher information contains a factor $\gamma_D\propto 1/M$. This important difference shows that suppressing vacuum term with the weak measurement is the reason for an extra improvement by a factor of $M$ in the local scheme with quantum randomness.

\section{Compare to the scheme of sending resources to all telescopes}\label{SI:sufficient_resource}

%We have been discussing the case that we just distribute resources to one pair of telescopes. In the classical randomness case, this means lots of stellar photons are lost. One may ask what if in the classical randomness we just send resources to all telescopes and do the measurement, how is this compared to the schemes using quantum randomness. The answer is that by wasting lots of resources to send resources to all telescopes, the performance will be exactly as the scheme using quantum randomness. It is equivalent to say that our scheme using quantum randomness can save the resource by a factor scaling as $M$ while having the same performance.

We have been considering the scenario where bipartite resources are distributed to only one pair of telescopes. In the case of classical randomness, this approach results in the loss of many stellar photons. One might ask what would happen if we instead distributed bipartite or multipartite resources to all telescopes, regardless of whether the telescope actually receives the single photon.  How does this compare to schemes that use quantum randomness, i.e. pre-processing of the stellar photon? The answer is that, not only is sending resources to all telescopes highly inefficient, but its performance still roughly matches that of the scheme using pre-processing. In other words, our quantum randomness approach achieves the same performance while saving on the distributed resources.

%For the Gottesman protocol, if we distribute terrestrial photons to all pairs of $M$ telescopes (assume $M$ is even for simplicity), note that there are $M(M-1)/2$ pairs of telescopes, each time we can just select one possible set of $M/2$ pairs of telescopes. This means to estimate all coherence functions $g_{XY}$, we need the POVM

\subsection{GJC protocol with bipartite entangled states}
If we distribute terrestrial photons to all pairs of $M$ telescopes (assuming $M$ is even for simplicity), note that there are $M(M-1)/2$ possible pairs of telescopes. Each time, we can only select one set of $M/2$ pairs of telescopes. This implies that, to estimate all coherence functions $g_{XY}$, we need to perform a POVM 
\begin{equation}
E_{XY}=\frac{1}{M-1}\frac{1}{2}\ket{\delta_\pm}\bra{\delta_\pm}_{XY}\otimes \ket{00\cdots }\bra{00\cdots}_{\overline{XY}}
\end{equation}
where the factor of $1/(M-1)$ arises from the $M-1$ possible ways of forming $M/2$ pairs of telescopes, and the factor of $1/2$ comes from the fact that the GJC protocol loses half of the stellar photons \cite{gottesman2012longer}. By directly comparing this with the POVM for the GJC scheme using quantum randomness in Eq.~\ref{eq:POVM_Gottesman_quantum_M}, where the optimal $\gamma_D = 1/(M-1)$, we conclude that both schemes should have the same performance. However, it is crucial to note that the GJC protocol with quantum randomness only requires sending a terrestrial photon to a single pair of telescopes, thereby saving the entanglement resource by a factor of $M/2$.

\subsection{GJC protocol with multipartite entangled states}
Instead of sending a bipartite entangled state to a pair of telescopes, suppose the central server distributes a multipartite entangled W state of the form
\begin{equation}
\ket{W}=\frac{1}{\sqrt{M}}(e^{i\delta_1}\ket{100\cdots 0}+e^{i\delta_2}\ket{010\cdots0}+\cdots+e^{i\delta_M}\ket{000\cdots 1}).
\end{equation}
A modified GJC protocol is performed on the state $\rho_s^{(1)} \otimes \ket{W}\bra{W}$ by interfering the stellar light and terrestrial light at each telescope and measuring the photon number in the two output ports.  At the $i$th telescope, there are three relevant measurement outcomes for our discussion:
(1) Both ports detect vacuum, projecting the state onto $\ket{0_{is}0_{it}}$, where subscript $i$ denotes the $i$th telescope, and $s,t$ refer to the stellar and terrestrial photons, respectively.
(2) One of the two output ports detects two photons, projecting the state onto $\ket{1_{is}1_{it}}$, which corresponds to the scenario where the $i$th telescope receives a single photon from both the stellar and terrestrial light.
(3) One output port detects a single photon, while the other port detects vacuum.  In this case, the $j$th telescope receives a single photon, forming a pair with the $i$th telescope. This outcome projects the state onto $\frac{1}{\sqrt{2}}(\ket{0_{is}1_{it}}\ket{1_{js}0_{jt}} \pm \ket{1_{is}0_{it}}\ket{0_{js}1_{jt}})$.
Since we cannot distinguish whether a single photon originates from stellar light or terrestrial light after interfering the stellar and terrestrial light, we now project onto a superposition of the two possibilities.

%We will do measurement on state $\rho_s^{(1)}\otimes\ket{W}\bra{W}$ by interfering the stellar light and terrestrial light at each telescope and measure photon number in the two output ports. At the $i$th telescope, this measurement has three different outcomes in the two output ports which are relevant for our discussion: (1) Both port get vacuum, which projects the state onto $\ket{0_{is}0_{it}}$, where subscript $i$ labels the $i$th telescope, $s,t$ label the stellar and terrestrial photons. (2) One of the two output ports gets two photons, which projects the state onto $\ket{1_{is}1_{it}}$ and corresponds to the case that $i$th telescope gets the single photon from both the stellar light and terrestrial light. (3) One of the output ports gets a single photn while the other output port gets vacuum. In this case, the $j$th telescope will get a single photon and form a pair with $i$th telescope. This measurement outcome project onto state $\frac{1}{\sqrt{2}}(\ket{0_{is}1_{it}}\ket{1_{js}0_{jt}}\pm\ket{1_{is}0_{it}}\ket{0_{js}1_{jt}})$.

We consider a $4\times 4$ block of the state $\rho_s^{(1)}\otimes\ket{W}\bra{W}$ for the telescope $i,j$,
\begin{equation}\begin{aligned}
\frac{1}{M}\left[
\begin{matrix}
1 & g_{ij}\\
g_{ij}^* & 1
\end{matrix}\right]\otimes \frac{1}{M}\left[
\begin{matrix}
1 & e^{i\delta_{ij}}\\
e^{-i\delta_{ij}} & 1
\end{matrix}\right]=
\frac{1}{M^2}\left[
\begin{matrix}
1 & e^{i\delta_{ij}} & g_{ij} & g_{ij}e^{i\delta_{ij}}\\
e^{-i\delta_{ij}} & 1 & g_{ij}e^{-\delta_{ij}} & g_{ij} \\
g_{ij}^* & g_{ij}^*e^{i\delta_{ij}} & 1 & e^{i\delta_{ij}}\\
g_{ij}^*e^{-i\delta_{ij}} & g_{ij}^* & e^{-i\delta_{ij}} & 1
\end{matrix}\right]
\end{aligned}\end{equation}
which is written in basis $\{\ket{0_{is}0_{it}}\ket{1_{js}1_{jt}},\ket{0_{is}1_{it}}\ket{1_{js}0_{jt}},\ket{1_{is}0_{it}}\ket{0_{js}1_{jt}},\ket{1_{is}1_{it}}\ket{0_{js}0_{jt}}\}$, $\delta_{ij}=\delta_i-\delta_j$. Note that when we write out all the $4 \times 4$ blocks for each pair of telescopes, the basis $\ket{1_{is}1_{it}}$ for each $i$ appears $M-1$ times, because all $M-1$ pairs of telescopes involving the $i$th telescope have a chance that the $i$th telescope receives two photons. As discussed above, the measurement will project the state onto $\frac{1}{\sqrt{2}}(\ket{0_{is}1_{it}}\ket{1_{js}0_{jt}} \pm \ket{1_{is}0_{it}}\ket{0_{js}1_{jt}})$, which occurs with a probability of
\begin{equation}
P_{ij}(\pm)=\frac{1}{M^2}[1\pm (\text{Re} g_{ij}\cos\delta_{ij} +\text{Im} g_{ij}\sin\delta_{ij})],\quad i>j
\end{equation}
The measurement of projecting onto $\ket{1_{is}1_{it}}\ket{0_{js}0_{jt}}$, (note $\ket{1_{is}1_{it}}\ket{0_{js}0_{jt}}=\ket{0_{1s}0_{1t}}\ket{0_{2s}0_{2t}}\cdots \ket{1_{is}1_{it}}\cdots \ket{0_{Ms}0_{Mt}}$), has probability
\begin{equation}
P_i=\frac{1}{M^2}
\end{equation}
It is straightforward to verify that $\sum_{i>j}(P_{ij}(+)+P_{ij}(-)) + \sum_i P_i = 1$, as expected. Now, by incorporating the fact that $\rho_s = (1-\epsilon)\rho^{(0)}_s + \epsilon\rho^{(1)}_s + o(\epsilon)$, we introduce a prefactor $\epsilon$ to the probability distribution above. With this in place, we can proceed to calculate the Fisher information
\begin{align}
    F_{XY}^W=\frac{2\epsilon}{M^2\left[1-Re(g_{XY}e^{-i\delta_{XY}})^2\right]}\begin{pmatrix}\cos^2\delta_{XY}&\sin\delta_{XY}\cos\delta_{XY}\\\sin\delta_{XY}\cos\delta_{XY}&\sin^2\delta_{XY}\end{pmatrix}
\end{align}
Compared to the case of GJC protocol using bipartite entangled state with quantum randomness case, ratio of Fisher information
\begin{equation}
\frac{\Vert F^{W}\Vert}{\Vert F^{qr}\Vert}=2\frac{M-1}{M}\rightarrow2,\quad M\rightarrow\infty
\end{equation}
As the number of telescopes $M \rightarrow \infty$, using the W state can improve the Fisher information of GJC protocol with quantum randomness by a factor of 2. Intuitively, this is because in the GJC protocol using bipartite entangled states, the protocol fails if both the terrestrial and stellar photons are detected at the same telescope. In the bipartite entanglement case, this failure occurs with a 50\% probability. However, when using $W$ states, this failure probability decreases to $\sum_i P_i=1/M$ as calculated above. It is important to note that this improvement from the W state is not due to our pre-processing step, but rather stems from the GJC protocol itself, which relies only on linear optics \cite{gottesman2012longer}. It is possible to avoid this loss of the factor of 1/2 by using nonlinear optical elements while still employing bipartite entanglement, as discussed in Ref.~\cite{czupryniak2023optimal}. This factor depends solely on the measurement conducted after the pre-processing steps outlined in our work. The key takeaway is that our pre-processing step does not result in the loss of stellar photons during the process of collapsing the state onto a single pair of telescopes. And collapsing the state onto a single pair of telescopes significantly simplifies the distribution of entanglement resources.

\subsection{Local scheme with multipartite shared reference frame}

For the local scheme in which we distribute the resources used for the reference frame to all telescopes, we will project onto $\ket{\delta_{j},\alpha_j}=(\ket{0}+ (-1)^{\alpha_j}e^{i\delta_j}\ket{1})/\sqrt{2}$, where $\alpha_j=0,1$ for the two basis states, $\ket{0},\ket{1}$ are the vacuum and single photon states at the spatial mode corresponding to the $j$th telescope. Note that this requires the central server to distribute resource states that are phase-locked across all telescopes, which constitutes a multipartite resource.
For the stellar state described in Eq.~\ref{rhoM}, the probability of getting outcome described by $\vec{\alpha}=[\alpha_1,\alpha_2,\cdots,\alpha_M]$, 
\begin{equation}
P(\vec{\alpha})=\frac{1}{2^M}+\frac{1}{2^{M-1}}\sum_{j_1>j_2}(-1)^{\alpha_{j_1}+\alpha_{j_2}}\frac{\epsilon}{M}(\text{Re}g_{j_1j_2}\cos\delta_{j_1j_2}+\text{Im}g_{j_1j_2}\sin\delta_{j_1j_2})
\end{equation}
where $M$ is the number of telescopes, Fisher information for the estimation of each coherence function $g_{XY}$ has the block form
\begin{align}
F_{XY}&=\sum_{\alpha}\frac{{\epsilon}^2/{(M2^{M-1})}^2}{P(\vec{\alpha})}\begin{pmatrix}\cos^2\delta_{XY}&\sin\delta_{XY}\cos\delta_{XY}\\\sin\delta_{XY}\cos\delta_{XY}&\sin^\delta_{XY}\end{pmatrix}\\
    &=\frac{4\epsilon^2}{M^2}\begin{pmatrix}\cos^2\delta_{XY}&\sin\delta_{XY}\cos\delta_{XY}\\\sin\delta_{XY}\cos\delta_{XY}&\sin^\delta_{XY}\end{pmatrix},
\end{align}
where we ignore the subleading order terms of $\epsilon$ in $P(\vec{\alpha})$ from the first row to the second row. By directly comparing this Fisher information to Eq.~\ref{eq:direct_FI}, we find that it achieves the same performance as the local scheme using quantum randomness. However, to implement the measurement described here, a shared phase reference frame must be established between all telescopes. %Therefore, the local scheme with quantum randomness offers a resource-saving advantage by a factor of $M/2$.

%Directly compare this Fisher information to Eq.~\ref{eq:direct_FI}, we find it has the same performance as the local scheme using quantum randomness. But to implement the measurement described here, we have to build a shared phase reference frame between all telescopes. So, local scheme using quantum randomness can save the resources by a factor of $M/2$.

In summary, the case where we distribute resources to all telescopes is essentially equivalent in performance to what we have presented in the main text. Quantum randomness can either be described as enhancing performance over classical randomness by a factor of $M/2$ with the same amount of resources, or alternatively, as simplifying resource distribution. In the latter case, instead of allocating resources to all telescopes, quantum randomness requires distributing  bipartite resource to just one pair of telescopes.

\end{document}